\documentclass[11pt]{article}
\usepackage{jheppub}
\usepackage[utf8]{inputenc}
\usepackage{amsmath}
\usepackage{graphicx}
\usepackage{xcolor}
\usepackage{comment}
\usepackage[toc,page]{appendix}
\usepackage{feynmp}
\usepackage{slashed}
\usepackage{tikz-feynman}
\tikzfeynmanset{compat=1.1.0}
\usetikzlibrary{positioning,arrows}
\usetikzlibrary{decorations.pathmorphing}
\usetikzlibrary{decorations.markings}

\graphicspath{{Images/}}

\title{One-point functions in 2D and 4D Janus interfaces}
\date{\today}
\author{Andreas Karch,}
\author{Ainesh Sanyal,}
\author{Ryan C. Spieler,}
\author{and Mianqi Wang}

\affiliation{${}^{1}$Theory Group, Weinberg Institute, Department of Physics,
University of Texas \\  2515 Speedway, Austin, TX 78712, USA.}

\emailAdd{karcha@utexas.edu}
\emailAdd{aineshsanyal@utexas.edu}
\emailAdd{rcspieler@utexas.edu}
\emailAdd{mqwang@utexas.edu}

\abstract
{We calculate one-point functions of the marginal operators dual to the space-varying dilaton in 4D and 2D holographic Janus interfaces, extending the results of \cite{Clark_2005}. We compare the strongly coupled supergravity and weakly coupled CFT limits for $\mathcal{N}=0,1,2,4$ Janus interfaces in 4D $\mathcal{N}=4$ SYM and for non-supersymmetric and half-BPS Janus interfaces in the 2D D1--D5 CFT. In 4D, the $\mathcal{N}=0,1$ results agree only to first order in the jump parameter,  and the $\mathcal{N}=2,4$ result agrees exactly. In 2D, exact agreement holds for the half-BPS interface, whereas the non-supersymmetric interface agrees only to first order.}

\begin{document}
\tikzset{
line/.style={thick, decorate, draw=black,}
 }
\maketitle

\section{Introduction}

Correlation functions of local operators have been studied extensively in the context of AdS/CFT \cite{Maldacena_1999,witten1998antisitterspaceholography,Freedman_1999,witten2002multitraceoperatorsboundaryconditions,Duff_2004,Aharony_2000}. In addition, the presence of defects or interfaces significantly enriches the correlation function structures and adds in new defect CFT (dCFT) or interface CFT (ICFT) data \cite{Bill__2016,andrei2018boundarydefectcftopen}. One of those additional data is the one point (1-pt) function $a_{O}$ of ambient operators in the CFT, which vanishes without the interfaces/defects. Via the gravity dual to ICFTs, correlation functions such as the one point function, bulk-to-defect correlators and bulk-to-bulk correlators have been calculated in various setups - see \cite{Nagasaki_2012,Nagasaki_2012e,Buhl_Mortensen_2016,deleeuw2017introductionintegrabilityonepointfunctions,de_Leeuw_2017,de_Leeuw_2020,Wang_2020,Komatsu_2020,He_2025} for examples. For recent reviews on this subject, see \cite{Linardopoulos_2020,linardopoulos2025stringtheorymethodsdefect}.

In this paper we focus on a specific family of holographic conformal interfaces in 2D and 4D CFTs called Janus interfaces. They are characterized by a jump in moduli of the CFTs at the interface. In the bulk the scalar fields dual to those moduli vary in spacetime and have specific boundary conditions near the two boundaries, via the AdS/CFT dictionary.

4D Janus was first given in \cite{Bak_2003} as a prototype of dilatonic deformation of $\mathcal{N}=4$ SYM/AdS$_5\times S^5$. The geometry of the bulk is reducible to 5D, featuring AdS$_4$ slicing of asymptotic AdS$_5$. It preserves the $SO(3,2)$ conformal symmetry, but completely breaks supersymmetry (SUSY). In \cite{Clark_2005}, the boundary of Janus is proposed to be a gauge theory with a space-varying gauge coupling $g_{YM}$. The operator $\mathcal{L}'$ dual to the dilaton is the fourth descendant of $\mathrm{tr}\,\phi^{\{i}\phi^{j\}}$, and differs from the standard gauge theory Lagrangian $\mathcal{L}$ by certain total derivatives, as we will review in the paper. 

Janus interfaces between $\mathcal{N}=4$ SYM were studied systematically in \cite{D_Hoker_2006}. It was argued that there existed $\mathcal{N}=4,2,1$ Janus interfaces preserving a portion of SUSY by adding interface Lagrangians governed by the BPS equations. Their holographic duals were studied since. In \cite{D_Hoker_2006janus,Clark_2005s} the 5D and 10D supergravity duals to the $\mathcal{N}=1$ interface with $SU(3)$ was solved, while in \cite{D_Hoker_2007} the maximal $\mathcal{N}=4$ Janus solution was solved. In \cite{Bobev_2020} the holographic duals to all three $\mathcal{N}=4,2,1$ interfaces were given using 5D gauged supergravity and 10D top-down supergravity.

In 2D the situation is slightly simpler. The CFT we focus on is the 2D $\mathcal{N}=(4,4)$ D1-D5 CFT dual to type IIB on AdS$_3\times S^3\times T^4$ in RR background. We also have good control of the free orbifold point on the conformal manifold. Janus interfaces in 2D were first given in \cite{Bak_2007} which broke all SUSY, and later in \cite{Chiodaroli_2010_half} the holographic dual of the half-BPS $\mathcal{N}=4$ interface was solved.

In 2D the marginal coupling that jumps across the interface is the $T^4$ sigma-model modulus, which is determined by the boundary value of the bulk dilaton. In addition, the $\mathbb{Z}_2$ twist operator near the orbifold point is dual to the axion and provides another Janus parameter, but we do not consider it here. At the free orbifold point the untwisted modulus has a simple current-current representative. For the half-BPS interface the relevant marginal operator contains both bosonic and fermionic terms, whereas the non-supersymmetric dilatonic Janus solution couples to the bosonic sigma-model modulus \cite{David_2002,Bak_2007}.

In \cite{Clark_2005} the one-point function of the dilaton operator $\mathcal{L}'$ in Janus without supersymmetry was calculated at strong and weak coupling and in conformal perturbation theory. The weak- and strong-coupling results agree only to first order in the gauge-coupling jump $\gamma$. We study the same question for supersymmetric interfaces in 4D and for Janus interfaces with and without supersymmetry in 2D.

Our results are as follows. For 4D $\mathcal{N}=1$ Janus, the weak- and strong-coupling one-point functions agree only to first order in $\gamma$, as in the non-supersymmetric case. For the $\mathcal{N}=2$ interface, numerical evaluation of the full bulk integral is consistent with an exact linear relation in the physical jump and hence with exact weak/strong matching; the maximally supersymmetric $\mathcal{N}=4$ interface also agrees exactly. In 2D, the non-supersymmetric Janus result agrees only to first order in the radius-squared (sigma-model-metric) jump, whereas the half-BPS result agrees exactly.

The $\mathcal{N}=2$ result shows that exact matching of this observable need not be restricted to half-BPS interfaces. Together with exact weak/strong results for $c_\text{eff}$ \cite{Gutperle_2016}, $c_{LR}$ \cite{baig2024transmissioncoefficientsuperjanussolution}, and $\log g$ \cite{Chiodaroli_2010,karch2025nonrenormalizationtheoremcaln44}, this motivates identifying the supersymmetric structures that protect particular interface observables.

It is worth noting that Janus interfaces are not the most general holographic conformal interfaces (domain walls) that have been worked out, especially in dimensions higher than 2. In particular, there are top-down realizations where localized degrees of freedom such as SCFTs living on the conformal interfaces \cite{Karch_2001,Karch_2022}. For $\mathcal{N}=4$ SYM, there is a huge family of D3/D5/NS5 defects $\mathcal{N}=4$ SYM \cite{Gaiotto_2009,Gaiotto_2009s} where certain 3D $\mathcal{N}=4$ theories sit on them, and their holographic duals were known \cite{D_Hoker_2007,D_Hoker_20072,Assel_2011}. In \cite{He_2025} the one-point function of local operators were calculated for such interfaces using supersymmetry localization, and our results for the empty $\mathcal{N}=4$ interface agree with theirs.

The paper is organized as follows: in section \ref{sec:review} we review the 1-pt function calculation for the non-SUSY interface in \cite{Clark_2005}. In section \ref{sec:4dgravity} we present the bulk calculation of the 1-pt function of the operator dual to dilaton for $\mathcal{N}=1,2,4$ interfaces in SYM. In section \ref{sec:4dcft} we conduct the calculation in the weakly-coupled limit and compare with the gravity results. Finally in section \ref{sec:2d} we calculate the 1-pt function in 2D Janus and SUSY Janus interfaces and compare the free orbifold results with the gravity ones.

\paragraph{Note Added:}
In an earlier version of this paper we failed to properly take into account the uplift of the dilaton to 10D. This was pointed out in \cite{christophtoappear}. In response, this revised version of the manuscript now reports much more interesting results, reporting agreement of 1-pt functions between strong and weak coupling not just only for 1/2 BPS interfaces, but also for 1/4 BPS interfaces in 4D.

\section{Review: 1-pt function in non-SUSY Janus of $\mathcal{N}=4$ SYM}
\label{sec:review}
Let us follow \cite{Clark_2005} and review the 1-pt function of the marginal operator dual to the dilaton for non-SUSY Janus interface in $\mathcal{N}=4$ SYM.

 The non-SUSY Janus solution is given by the dilatonic deformation of type IIB on AdS$_5\times S^5$ and breaks all of the supersymmetry. The geometry is AdS$_4\times\mathbb{R}\times S^5$ with a warpfactor $A(\mu)$ on the AdS$_4$ slices. We can solve for the metric with the warpfactor, dilaton $\phi(\mu)$ and 5-form RR field strength \cite{Bak_2003}. The asymptotic form of the dilaton near the boundary $\mu\to\pm\mu_0$ is given by
\begin{equation}
    \phi(\mu)=\phi_\pm\mp \frac{c}{4}e^{-4A(\mu)}+\dots,
\end{equation}
where the asymptotic values of $\phi_\pm$ near the boundary are related to the dimensionless constant $c$ by
\begin{equation}
    \phi_{\pm} = \phi_0 \pm \frac{c\sqrt{\pi}}{2} \sum_{n=0}^{\infty}
\frac{\Gamma(4n + 2)}{\Gamma\!\left(3n + \frac{5}{2}\right)\, n!} \, b^n
= \phi_0 \pm \frac{2}{3}c + \mathcal{O}(c^3),
\label{eq:candgamma}
\end{equation}
for some arbitrary constant $\phi_0$ and $b=c^2L^2/24$ with AdS radius $L$.

Here, the warpfactor scales as $A(\mu)\sim 1/\sin(\mu_0-|\mu|)$ near the boundary and $f=ze^{-A}$ serves as the Fefferman-Graham (F-G) parameter. We want to calculate the expansion of the dilaton near the boundary. To the second order we have the vacuum expectation of the operator dual to the dilaton \cite{Aharony_2000,Clark_2005}
\begin{equation}
    e^{\phi(\mu)}=e^{\phi_\pm}\left(1-\frac{2\pi^2}{N^2}\langle\mathcal{L'}\rangle z^4 e^{-4A(\mu)}+\dots\right).
    \label{eq:lagrangian}
\end{equation}

Throughout the paper we use the standard convention for the dilaton, string coupling and Yang-Mills coupling:
\begin{equation}
    e^{\phi}=g_s=\frac{g_{YM}^2}{4\pi}.
    \label{eq:dilaton}
\end{equation}

Here the proposed gauge theory dual to the Janus involves a variant of the original Lagrangian $\mathcal{L}'$ that is different from the ordinary Lagrangian $\mathcal{L}$ by a total derivative term $\partial(X^I\partial X^I)/2$ \cite{Clark_2005}.

The argument for \eqref{eq:lagrangian} is as follows: 
From the AdS$_5$/CFT$_4$ dictionary, if an action is deformed by a massless primary field with dimension $\Delta=d=4$ coupled to a source, the expansion of the bulk field $\phi_\Delta$ near the boundary is 
\begin{equation}
     \phi_{\Delta=4}\sim a_f(1+\dots)+b_f(f^4+\dots).
 \end{equation}
 
 In the dual field theory, the coupling of the field is $a_f$, while the VEV of the field is $\langle O_{\Delta=4}\rangle=-4b_f$. The dual operator $O$ is related to $\mathcal{L}'$ by a factor of $1/16\pi G_5$ \cite{Balasubramanian_1999,Aharony_2000}. Using the relation between the Newton's constant and large N factor under unit radius $G_5=\pi/2N^2$, we derived the subleading piece in \eqref{eq:lagrangian}.

In the above non-SUSY Janus solution the 1-pt function of $\mathcal{L}'$ is given by (interface located at $x^3=0$)
\begin{equation}    
\langle\mathcal{L}'\rangle=\frac{N^2c}{8\pi^2(x^3)^4}\varepsilon(x^3)
\label{eq:nonsusy4d}
\end{equation}

The CFT dual to the space-varying dilaton configuration above is proposed to be a gauge theory with $g_{YM}(x)$ jumping across the interface \cite{Clark_2005}. We assume the interface sits at $x^3=0$ throughout the paper ($x^1=0$ for the 2D cases). Concretely, it is given by the action 
\begin{equation}
    S=\int d^4x\left(\mathcal{L}'_{\rm ref}-\gamma\,\varepsilon(x^3)\mathcal{L}'_{\rm ref}\right)
\end{equation}

where $\mathcal{L}'_{\rm ref}=\bar g^{-2}\mathcal{L}_0'$ and $\bar g^{-2}=\frac{1}{2}(g_+^{-2}+g_-^{-2})$. The density $\mathcal L'_0$ differs from the usual SYM density by $\frac12\partial_\mu\mathrm{tr}(\phi^iD^\mu\phi^i)$. The local dilaton operator is $\mathcal L'=g(x^3)^{-2}\mathcal L'_0=(1-\gamma\varepsilon(x^3))\mathcal L'_{\rm ref}$, with $\mathcal L'=-\delta S/\delta\phi$ at fixed fields. The jump parameter $\gamma$ is related to \eqref{eq:candgamma} by
\begin{equation}
    \gamma\equiv\frac{g_+^2-g_-^2}{g_+^2+g_-^2}=\tanh\left(\frac{\phi_+-\phi_-}{2}\right)=\frac{2}{3}c+\frac{16}{189}c^3+O(c^5).
    \label{eq:gamma}
\end{equation}

Following \cite{Clark_2005}, the D3 probe action in Einstein frame motivates the weak-coupling description. The dilaton couples directly to the gauge flux, while the scalars have canonical kinetic terms. Writing $\Phi^i=\phi^i/g$, $i=1,\ldots,6$, and using Majorana fermions $\lambda^A$, $A=1,\ldots,4$, the Lagrangian is
\begin{equation}
\begin{split}
\mathcal{L} = \mathrm{Tr} \Big[
& -\frac{1}{2} D_\mu \Phi^i D^\mu \Phi^i
- \frac{1}{2} i \bar{\lambda}^{A} \slashed{D} \lambda^{A}
- \frac{1}{4 g^2(x^3)} F_{\mu\nu}^2 \\
& + \frac{g(x^3)}{2} \bar{\lambda} \Gamma_i [\Phi^i, \lambda]
+ \frac{g(x^3)^2}{4} [\Phi^i, \Phi^j]^2
\Big].
\end{split}
\label{eq:probe-lagrangian}
\end{equation}

Here $\Gamma_i$ are the internal Clifford matrices in the Yukawa coupling, and $\mathrm{Tr}$ is the gauge trace. One can check that the kinetic terms of the above $\mathcal{L}$ recover the desired marginal operator $\mathcal{L}'(\phi^i,\psi^\alpha)$ dual to the dilaton. The total derivative term $\phi^i\Box\phi^i/g^2$ comes from the kinetic term of $\Phi$, whereas the Majorana fermions do not contain extra terms after the rescaling. Indeed, $\mathcal{L}'$ is the correct on-shell Lagrangian for $\mathcal{N}=4$ SYM, which is the fourth descendant of $\mathrm{tr}\,\phi^{\{i}\phi^{j\}}$ in the stress tensor supermultiplet \cite{Clark_2005,Chicherin_2017}. We will go back to this in Section \ref{sec:4dcft} where we determine the exact boson and fermion kinetic terms in $\mathcal{L}'$ for more general interfaces.

In the weakly-coupled regime with small $g_{YM}$, the 1-pt function of the Lagrangian is given by the (derivatives of) tree-level free propagators of the boson and fermion fields at the contact limit $x=y$:
\begin{equation}
\langle \mathcal{L}'(x) \rangle
= \mathrm{Tr} \Big[
\frac{1}{2} \langle \partial^\mu \Phi^i(x)\, \partial_\mu \Phi^i(x) \rangle
+ \frac{1}{2} \langle \bar{\lambda}^A(x)\, \slashed{\partial} \lambda^A(x) \rangle
+ \frac{1}{4 g(x^3)^2} \langle F_{\mu\nu}(x)\, F_{\mu\nu}(x) \rangle
\Big].
\label{eq:1-ptcft}
\end{equation}

Using the method of image charges, only the gauge-field propagator contributes a finite term. The scalar and fermion propagators are the ordinary vacuum propagators in the canonical fields, so their contributions are vacuum coincidence singularities, removed by vacuum subtraction. The gauge-field matching conditions determine $G_{\mu\nu}=\langle A_{\mu}(x)A_{\nu}(y)\rangle$, from which we extract the image part of the field-strength correlator,
\begin{equation}
\langle F_{\mu\nu}(x)\, F_{\rho\sigma}(y) \rangle_{\text{image}}
= \frac{1}{\pi^2}
\frac{
J_{\mu\tau}(x - Ry)\, J_{\nu\lambda}(x - Ry)
- (\tau \leftrightarrow \lambda)
}{
(x - Ry)^4
}
R_{\,\,\rho}^\tau\, R_{\,\,\sigma}^\lambda.
\end{equation}

where $R_{\mu\nu}=\mathrm{diag}(1,1,1,-1)$ and $J_{\mu\nu}(x)=\delta_{\mu\nu}-\frac{2x_\mu x_\nu}{x^2}$. The result for \eqref{eq:1-ptcft} is exact to all orders of the jump $\gamma$ but only to leading order of $\bar{g}$ \cite{Clark_2005}:
\begin{equation}
    \langle\mathcal{L}'\rangle=\gamma \frac{3N^2}{16\pi^2(x^3)^4}\varepsilon(x^3)
    \label{eq:cft4d}
\end{equation}

Comparing \eqref{eq:nonsusy4d} and \eqref{eq:cft4d}, we find that for the non-SUSY Janus interface, the 1-pt function of this variant of the Lagrangian $\mathcal{L}'$ only matches in the weakly- and strongly-coupled sides to the leading order of the jump $\gamma$ (or $c$). Explicitly, the gravity result is
\begin{equation}    \langle\mathcal{L}'\rangle=\frac{3N^2}{16\pi^2(x^3)^4}\left(\gamma-\frac{2}{7}\gamma^3+\dots\right)\varepsilon(x^3)
\label{eq:nonsusygrav}
\end{equation}

\section{Gravity duals of holographic SUSY interfaces in 4D}
\label{sec:4dgravity}
As mentioned above, there are holographic SUSY Janus solutions dual to $\mathcal{N}=4,2,1$ interfaces in SYM. In this section we follow the supergravity solutions of \cite{Bobev_2020} and calculate the one-point functions determined by their dilaton profiles. The $\mathcal{N}=1$ integral can be evaluated numerically and expanded analytically at small jump. For $\mathcal{N}=2$, numerical evaluation of the full integral is consistent with a simple closed relation, while the $\mathcal{N}=4$ solution is analytic.

\subsection{$\mathcal{N}=1$ interface}
The 10D supersymmetric Janus solution with an $\mathcal{N}=1$ interface was introduced in \cite{D_Hoker_2006janus} and section 4 of \cite{Bobev_2020}. The flavor symmetry is $SU(3)$. The geometry is AdS$_4\times CP^2\times_1 S^1\times\mathbb{R}$. It is convenient to study the solution in terms of the 5D Janus model embedded in the maximal gauged $SO(6)$ supergravity.

We follow the notations in \cite{Bobev_2020}. The metric in terms of the slicing coordinate $r$ is
\begin{equation}
    ds_5^2=dr^2+e^{2A(r)}ds_{AdS_4}^2
    \label{eq:janus}
\end{equation}

The solution to the BPS equations is as follows: the warpfactor is given by the constant $0<\mathcal{I}\le1$ and coordinate $X$,
\begin{equation}
    e^{2A}=\frac{5^{5/3}\,\mathcal{I}}{9g^2}e^{2X}
\end{equation}

and $X$ is related to $r$ by
\begin{equation}
    \frac{4}{g^2}\left(\frac{dX}{dr}\right)^2+V_{\text{eff}}=0
\end{equation}

where
\begin{equation}
    V_\text{eff}=4e^{-2X}\left(\frac{9}{5^{5/3}\mathcal{I}}-e^{-4X}\cosh^23X\right)
\end{equation}

Below we use $g=2/L=2$, where $L$ is the AdS radius, and restrict to the axion-free representative with $\mathcal{J}=c_0=0$. The coordinate $X$ approaches $\infty$ in both asymptotic AdS$_5$ regions and attains its minimum $X_{\mathrm{tp}}$ where $V_\text{eff}$ vanishes. Denoting the five-dimensional scalar of \cite{Bobev_2020} by $\varphi$, its solution is
\begin{equation}    \varphi(X)=\varphi_0\pm\int_{X_{\mathrm{tp}}}^X\frac{9\,e^{-x}}{5^{5/6}\sqrt{\mathcal{I}}\cosh 3x}\frac{dx}{\sqrt{-V_\text{eff}(x)}}
\label{eq:jump}
\end{equation}

Recall that, in terms of $X$, the metric is
\begin{equation}    ds^2=\frac{5^{5/3}\,\mathcal{I}}{36}e^{2X}ds_{AdS_4}^2+\frac{dX^2}{-V_\text{eff}}    
\end{equation}

We extract the subleading coefficient in terms of the Fefferman--Graham parameter
\begin{equation}
 f^2=z^2e^{-2A}=\frac{36z^2}{5^{5/3}\mathcal{I}}e^{-2X}.
\end{equation}
It is essential not to identify $\varphi$ directly with the standard type-IIB dilaton $\Phi$ appearing in \eqref{eq:dilaton}. For the axion-free uplift, $\Phi=\pm2\varphi$ where the sign is fixed by the choice of interface orientation \cite{Bobev_2020}. We set
\begin{equation}
 \eta_5\equiv\frac{|\varphi_+-\varphi_-|}{2},
 \label{eq:eta5}
\end{equation}

so the jump parameter in the coupling constants is

\begin{equation}
 \gamma\equiv \left|\frac{g_+^2-g_-^2}{g_+^2+g_-^2}\right|
 =\tanh\!\left(\frac{|\Phi_+-\Phi_-|}{2}\right)=\tanh(2\eta_5),
\end{equation}

where we choose the side labels so that $\gamma\geq0$.

Expanding the physical dilaton near the asymptotic regions gives
\begin{equation}
    \Phi(X)=\Phi_\pm\mp\frac{9e^{-4X}}{5^{5/6}\sqrt{\mathcal{I}}}+\dots
    =\Phi_\pm\mp\frac{5^{5/2}\mathcal{I}^{3/2}}{144z^4}f^4+\dots .
\end{equation}

Although the full integral is difficult to evaluate analytically, its small-$\mathcal{I}$ expansion can be obtained from \eqref{eq:jump} between $X_{\mathrm{tp}}$ and $\infty$. With $A=5^{5/2}/108$, one finds

\begin{equation}
 \eta_5=A\,\mathcal{I}^{3/2}+\frac{128}{105}A^3\,\mathcal{I}^{9/2}
 +O(\mathcal{I}^{15/2}).
\end{equation}
Equivalently, if $y=2A\,\mathcal{I}^{3/2}=5^{5/2}\,\mathcal{I}^{3/2}/54$, then
\begin{equation}
 \gamma=\tanh(2\eta_5)=y-\frac{y^3}{35}+O(y^5).
\end{equation}

Hence, the one-point function of $\mathcal{L}'$ and its small-jump expansion are
\begin{equation}    \langle\mathcal{L}'\rangle=\frac{5^{5/2}N^2\mathcal{I}^{3/2}}{288\pi^2z^4}\varepsilon(x^3)
=\frac{3N^2}{16\pi^2z^4}\left(\gamma+\frac{\gamma^3}{35}+O(\gamma^5)\right)\varepsilon(x^3).
\label{eq:n=1}
\end{equation}

\subsection{$\mathcal{N}=2$ interface}

The 5D supergravity Janus solution for $\mathcal{N}=2$ interface and its uplift to 10D was given in section 3 of \cite{Bobev_2020}. The R-symmetry is $U(1)$ and the flavor symmetry is $SU(2)$. Let us study the 5D gauged supergravity and extract the dilaton profile. The metric is the same as \eqref{eq:janus}, where we substitute the warpfactor with
\begin{equation}
    e^{-2A}=\frac{2\sqrt{X}}{\sqrt{\mathcal{I}}}
\end{equation}

The BPS equation reads
\begin{equation}
    \left(\frac{dX}{dr}\right)^2+V_{\text{eff}}=0,
\end{equation}

where 
\begin{equation}
    V_{\text{eff}}(X)=-\frac{16(1-X)^{1/3}X^2}{\sqrt{\mathcal{I}}}\left(\sqrt{\mathcal{I}}-2\sqrt{X(1-X)}\right).
\end{equation}

The asymptotic AdS$_5$ regions are both at $X\to 0$, and $X$ reaches its maximal value at the turning point $X_{\mathrm{tp}}=\tfrac{1}{2}(1-\sqrt{1-\mathcal{I}})$. The slicing coordinate $r$ is related to $X$ by
\begin{equation}
    r(X)=\pm\int_{X_{\mathrm{tp}}}^X\,\frac{\,dx}{\sqrt{-V_{\text{eff}}(x)}}.
\end{equation}

For the axion-free representative with $\mathcal{J}=c_0=\psi_0=0$, the five-dimensional scalar $\varphi$ is given by
\begin{equation}    \varphi(X)=\varphi_0\pm\int_{X_{\mathrm{tp}}}^X\frac{(3-2x)x}{\sqrt{\mathcal{I}}\,(1-x)^{4/3}+2(1-x)^{5/6}\,x^{3/2}}\frac{dx}{\sqrt{-V_{\text{eff}}(x)}}.
\label{eq:dilaton5d}
\end{equation}

As above, the Fefferman--Graham parameter is
\begin{equation}
 f^2=z^2e^{-2A}=\frac{2z^2\sqrt{X}}{\sqrt{\mathcal{I}}}.
\end{equation}
In terms of $X$, the metric is
\begin{equation}    ds^2=\frac{\sqrt{\mathcal{I}}}{2\sqrt{X}}ds_{AdS_4}^2+\frac{dX^2}{-V_\text{eff}}.    
\end{equation}

The ten-dimensional uplift, followed by the constant $SL(2,\mathbb{R})$ rotation to the axion-free duality frame, gives \cite{Bobev_2020}
\begin{equation}
 \Phi=2\varphi+O(X^{3/2}),\qquad \Phi_\pm=2\varphi_\pm.
\end{equation}
Since $X=O(f^4)$, the correction does not affect the normalizable $f^4$ coefficient. The physical dilaton therefore has the asymptotic expansion
\begin{equation}
    \Phi(r\to\pm\infty)=\Phi_\pm\mp \frac{3}{2\sqrt{\mathcal{I}}}X+\dots
    =\Phi_\pm \mp \frac{3\sqrt{\mathcal{I}}}{8z^4}\,f^4+\dots .
\end{equation}

From \eqref{eq:dilaton5d}, the jump in the 5D dilaton value defined in \eqref{eq:eta5} is
\begin{equation}
 \eta_5
 =\frac{\sqrt{\mathcal{I}}}{2}+\frac{\mathcal{I}^{3/2}}{6}+O(\mathcal{I}^{5/2}).
\end{equation}

We can also set
$t=2\sqrt{x(1-x)}/\sqrt{\mathcal{I}}$ and then expand the integrand in \eqref{eq:dilaton5d} numerically. Up to machine precision, the jump parameter has a simple expression\footnote{We would like to thank Christoph Uhlemann for pointing this out.}
\begin{equation}
    \eta_5=\frac{1}{2}\mathrm{arctanh}\sqrt{\mathcal{I}}.
\end{equation}

This implies
\begin{equation}
 \gamma=\tanh\!\left(\frac{|\Phi_+-\Phi_-|}{2}\right)
 =\tanh(2\eta_5)=\sqrt{\mathcal{I}}.
\end{equation}

Using \eqref{eq:lagrangian}, the 1-pt function is numerically evaluated to be
\begin{equation}    \langle\mathcal{L}'\rangle
=\frac{3N^2\gamma}{16\pi^2z^4}\,\varepsilon(x^3).
\label{eq:n=2}
\end{equation}

\subsection{Maximal $\mathcal{N}=4$ interface}
In this subsection we follow the notation of \cite{D_Hoker_2007} for the uplifted 10D half-BPS Janus solution and study the dilaton expansion. The geometry is AdS$_4\times S^2\times S^2\times\Sigma$. The R-symmetry is $SU(2)\times SU(2)$. 

There is an important notational caveat that the dilaton in \cite{D_Hoker_2007} is off by a factor of 2 compared to the standard string theory dilaton in \eqref{eq:dilaton}. Namely, the $\phi$ in this section has $e^{2\phi}=g_s=g_{YM}^2/4\pi$. We have an analytic expression for $\phi(x,y)$ in terms of the non-compact $x$ and compact $y$. At $x\to\pm\infty$ the geometry factors and we get the asymptotic values $e^{2\phi_\pm}=g_{\pm}^2/4\pi$. Setting $r=e^{2\phi_+-2\phi_-}=g_+^2/g_-^2$, we have:

\begin{equation}   
e^{4\phi(x,y)}=e^{2\phi_++2\phi_-}\left(\frac{re^x+re^{-x}}{e^x+re^{-x}}\right)\frac{N(x,y)}{D(x,y)},
\end{equation}
where 
\begin{equation}
    \begin{split}
        N(x,y)=e^{3x} + r e^{-3x} + (1 + 2 r) e^x + (2 + r) e^{-x} - (1 - r) (e^x - e^{-x}) \cos 2 y\\
        D(x,y)=e^{3x} + r^2 e^{-3x} + r(1 + 2 r) e^{-x} + (2 + r) e^{x} - (1 - r) (e^x - re^{-x}) \cos 2 y\\.
    \end{split}
\end{equation}

 Near the boundary, the expansion is
\begin{equation}
    \phi(x\to\infty,y)=\phi_+ -\frac{3}{4}\left(r^2-1\right) e^{-4x}+\dots,\quad \phi(x\to-\infty,y)=\phi_- -\frac{3}{4}\left(r^{-2}-1\right) e^{4x}+\dots .
\end{equation}

Near the boundary $x\to\pm\infty$, the 10D asymptotic metric is
\begin{equation}
\begin{split}
ds^2 \;\sim\; \frac{1}{z^2 \mu^2}
\Bigg(
& z^2\, d\mu^2
+ \frac{dx_1^2 + dx_2^2 - dt^2 + dz^2}{2(1+r^{\pm 1})} \\
& \qquad
+ z^2 \mu^2 \Big(
dy^2
+ \sin^2 y \, ds_{S_1^2}^2
+ \cos^2 y \, ds_{S_2^2}^2
\Big)
\Bigg)
+ \mathcal{O}(\mu^2),
\end{split}
\end{equation}

where $\mu=e^{\mp x}$. The dilaton expansion in terms of the Fefferman-Graham parameter $f_\pm=\sqrt{2(1+r^{\pm 1})}\,ze^{\mp x}$ near the boundary is
\begin{equation}
    \phi=\phi_\pm-\frac{3}{16z^4}\frac{r-1}{r+1}f_+^{4}\,\theta(x^3)-\frac{3}{16z^4}\frac{r^{-1}-1}{r^{-1}+1}f_-^{4}\,\theta(-x^3)+\dots. 
\end{equation}

Now we read off the VEV of $\mathcal{L}'$ from \eqref{eq:lagrangian}. Note that since the dilaton here is one half of the normal dilaton in \eqref{eq:dilaton}, the coefficient of the subleading term in the expansion is $-\frac{\pi^2}{N^2}\langle \mathcal{L}'\rangle$. Hence,
\begin{equation}
    \langle\mathcal{L}'\rangle=\frac{3N^2}{16\pi^2z^4}\frac{r-1}{r+1}\varepsilon(x^3)=\frac{3N^2}{16\pi^2z^4}\frac{g_+^2-g_-^2}{g_+^2+g_-^2}\varepsilon(x^3).
    \label{eq:n=4}
\end{equation}
Thus the expectation value of ${\cal L}'$ truncates at linear order in the coupling jump. As in the $\mathcal{N}=2$ case, this yields agreement with the weakly coupled field-theory result, which is linear in the jump.

\section{CFT results for $\mathcal{N}=1,2,4$ interfaces}
\label{sec:4dcft}
The gravity calculations above show that, to first order in the jump $\gamma$, the one-point function of the operator $\mathcal{L}'$ is the same for the non-supersymmetric and $\mathcal{N}=1,2,4$ interfaces. This is consistent with conformal perturbation theory, which protects the leading term and requires it to match the weak-coupling result. Beyond leading order, the gravity answers distinguish the cases: the $\mathcal{N}=0,1$ solutions have cubic corrections, whereas the $\mathcal{N}=2,4$ answers are linear in the physical jump (with the $\mathcal{N}=2$ statement supported by the numerical integral relation above).

In this section we clarify that $\mathcal{L}'$ stays the same as in section \ref{sec:review} even in the presence of supersymmetric interfaces, and show that the change of fermion boundary conditions by the additional interface terms does not affect its 1-pt functions for $\mathcal{N}=1,2,4$ interfaces.

Concretely, we follow \cite{D_Hoker_2006} and denote the six adjoint scalars by $\phi^i$, $i=1,\ldots,6$, and the four Weyl fermions by $\psi^A$, $A=1,\ldots,4$. We write $\Psi_L^A=(\psi^A,0)^t$ for their four-component embeddings. The matrices $\gamma^\mu$ and $C$ are the Dirac and charge-conjugation matrices, and $\rho^i$ are the antisymmetric $SU(4)$ Clebsch--Gordan matrices coupling two fermions to a scalar. Repeated field indices are summed, and $\operatorname{tr}$ denotes the gauge trace. The ambient SYM density is
\begin{equation}
\begin{split}
    \mathcal{L}_{SYM} &= -\frac{1}{4g^2} \operatorname{tr}\!\left(F^{\mu\nu} F_{\mu\nu}\right)
- \frac{1}{2g^2} \operatorname{tr}\!\left(D^\mu \phi^i D_\mu \phi^i\right)
+ \frac{1}{4g^2} \operatorname{tr}\!\left([\phi^i,\phi^j][\phi^i,\phi^j]\right)
\\[6pt]
&\quad - \frac{i}{2g^2} \operatorname{tr}\!\left(\bar{\Psi}_L\gamma^\mu D_\mu \Psi_L\right)
+ \frac{i}{2g^2} \operatorname{tr}\!\left(D_\mu \bar{\Psi}_L\,\gamma^\mu \Psi_L\right)
\\[6pt]
&\quad + \frac{1}{2g^2} \operatorname{tr}\!\left(
\Psi_L^t C \rho^i [\phi^i,\Psi_L]
+ \Psi_L^\dagger C (\rho^i)^* [\phi^i,\Psi_L^*]
\right).
\end{split}
\label{eq:symlag}
\end{equation}

Notice that the fermion kinetic term is symmetrized. It is related to the standard kinetic term $-\frac{i}{g^2}\bar{\psi}\bar{\sigma}^\mu D_\mu\psi$ by a total derivative term. Usually this has no effect on the physics, but as we will see below, the fermion terms in $\mathcal{L}'$ has precisely this form. In Weyl notation we use $\sigma^\mu=(\mathbf1,\vec\sigma)$ and $\bar\sigma^\mu=(-\mathbf1,\vec\sigma)$, where $\vec\sigma$ are the Pauli matrices, with metric $\operatorname{diag}(-1,1,1,1)$. The kinetic term differs from $-i\bar\psi^A\bar\sigma^\mu D_\mu\psi^A/g^2$ by a total derivative at constant coupling. For varying $g$, derivatives of the coupling enter the equations of motion and must be retained. The interface interactions are
\begin{equation}    
\begin{split}
    \mathcal{L}_I^{\mathcal{N}=4}&=
\frac{\partial_3 g}{g^3}
\operatorname{tr}
\left(
\frac{i}{2}\,\Psi_L^t C \Psi_L
+
\frac{i}{2}\,\Psi_L^\dagger C \Psi_L^*
-
4 i g^6 \tilde{\phi}^{\,2}
\left[
\tilde{\phi}^{\,4}, \tilde{\phi}^{\,6}
\right]
\right)\\
    \mathcal{L}_I^{\mathcal{N}=2}&=
\frac{\partial_3 g}{g^3} \operatorname{tr}\!\Big(
\;\frac{i}{2}\,\Psi_L^t C D^{(2)} \Psi_L
+ \frac{i}{2}\,\Psi_L^\dagger C D^{(2)} \Psi_L^* + g^4\, 2i\, \tilde{\phi}^2 [\tilde{\phi}^3, \tilde{\phi}^5]
- g^4\, 2i\, \tilde{\phi}^2 [\tilde{\phi}^4, \tilde{\phi}^6]
\Big)\\
    \mathcal{L}_I^{\mathcal{N}=1}
&= \frac{\partial_3 g}{g^3}\,\operatorname{tr}\!\left(
\frac{i}{2}\,\Psi_L^t C D^{(1)} \Psi_L
+ \frac{i}{2}\,\Psi_L^\dagger C D^{(1)} \Psi_L^*
\right)
\\[6pt]
&\quad + (\partial_3 g)\,\operatorname{tr}\!\left(
i\,\tilde{\phi}^1 [\tilde{\phi}^3,\tilde{\phi}^6]
+ i\,\tilde{\phi}^1 [\tilde{\phi}^4,\tilde{\phi}^5]
+ i\,\tilde{\phi}^2 [\tilde{\phi}^3,\tilde{\phi}^5]
- i\,\tilde{\phi}^2 [\tilde{\phi}^4,\tilde{\phi}^6]
\right)\\
\end{split}
\label{eq:interfaceL}
\end{equation}

Here $D^{(1)}=\operatorname{diag}(1,0,0,0)$ and $D^{(2)}=\operatorname{diag}(1,1,0,0)$ select one and two of the four fermion species; the $\mathcal N=4$ bilinear couples to all four. The scalar variables in \eqref{eq:interfaceL} follow the rescalings of \cite{D_Hoker_2006}: $\tilde\phi^{2,4,6}=\phi^{2,4,6}/g^2$ for $\mathcal N=4$; $\tilde\phi^2=\phi^2/g^2$ and $\tilde\phi^{3,4,5,6}=\phi^{3,4,5,6}/g$ for $\mathcal N=2$; and $\tilde\phi^i=\phi^i/g$ for $\mathcal N=1$. These expressions use the corresponding rewriting of the bulk scalar terms, while \eqref{eq:symlag} gives the ambient density in the original fields. The fermions are not rescaled in \eqref{eq:interfaceL}.

\subsection{Dual operator of space varying dilaton}
\label{sec:dilaton-operator}

Recall from section \ref{sec:review} that the dilaton couples to a fourth descendant in the $\mathcal N=4$ stress-tensor multiplet. Here let us be more precise and fix its lagrangian expression. For a real dilaton, the operator is the real combination of the chiral and anti-chiral descendants. It is represented by
\begin{equation}
 \mathcal{L}_{0SYM}'=\mathcal{L}_{0SYM}
 +\frac{1}{2}\partial_\mu\mathrm{tr}(\phi^iD^\mu\phi^i),
    \label{eq:4dld}
\end{equation}
where $\mathcal L_{0SYM}=g^2\mathcal L_{SYM}$ and $\mathcal L_{0SYM}'=\mathcal L'_0$ from \eqref{eq:symlag} and the notations in Section \ref{sec:review}. The scalars has the same unconventional form as in \cite{Clark_2005} and the fermion kinetic term is symmetrized.

To see this, consider the Abelian coupling operator $O_\tau$ in \cite{Basu_2004} sourced by the complexified coupling $\tau=\tau_1+i\tau_2$, with the theta term
$\tau_1=\theta_{YM}/(2\pi)$ and the dilaton
$\tau_2=4\pi/g^2=e^{-\phi}$.
In Abelian SYM theory, in the notation of \cite{Basu_2004} we have
\[
 O_\tau=\frac{\tau_2}{8\pi}\left[
 F_{\mu\nu}F^{\mu\nu}
 +iF_{\mu\nu}\widetilde F^{\mu\nu}
 -4i\bar\lambda_A\bar\sigma^\mu\partial_\mu\lambda^A
 +4\bar\varphi_{AB}\Box\varphi^{AB}\right],
 \qquad O_{\bar\tau}=O_\tau^\dagger .
\]
Varying the dilaton at fixed fields
and axion gives the combination operator
\[
 \delta S=\frac{i}{4\tau_2}\int d^4x\,
 (\delta\tau\,O_\tau-\delta\bar\tau\,O_{\bar\tau})
 =\frac14\int d^4x\,\delta\phi\,
 (O_\tau+O_{\bar\tau}).
\]

 The fermion term in the operator $\mathcal{L}'\sim O_\tau+O_{\bar{\tau}}$ is then proportional
to $-i(\bar\lambda\bar\sigma^\mu\partial_\mu\lambda-\partial_\mu\bar\lambda\bar\sigma^\mu\lambda)$, which reproduces
the symmetrized term for the real dilaton, with the
overall sign fixed as in section \ref{sec:review}.

The non-Abelian completion then follows from the descendant construction in \cite{Chicherin_2017} and the Appendix of \cite{Eden_2013}. Starting with the SYM density, its chiral descendant is obtained by adding the scalar derivative in \eqref{eq:4dld}, a fermion-current derivative and a gauge topological term, without using the EOM. The latter two additions cancel against their conjugates in the real combination, while the scalar derivative survives. This recovers \eqref{eq:4dld} with the SYM interactions unchanged.

Written in terms of the space-varying gauge coupling, the action is
\begin{equation}
    S=\int d^4x\frac{1}{g(x)^2}\mathcal{L}_{0SYM}'(x),
\end{equation}

with the full operator dual to the dilaton
\begin{equation}
 \begin{split}
 \mathcal L'(x)
 &=\frac{1}{g(x^3)^2}\mathcal L'_{0SYM}(x)\\
 &=\frac{1}{g(x^3)^2}\operatorname{tr}\!\Big[
 -\frac14F_{\mu\nu}F^{\mu\nu}
 +\frac12\phi^iD_\mu D^\mu\phi^i\\
 &\qquad-\frac{i}{2}\left\{
 \bar\psi^A\bar\sigma^\mu D_\mu\psi^A
 -(D_\mu\bar\psi^A)\bar\sigma^\mu\psi^A\right\}
 +\frac14[\phi^i,\phi^j][\phi^i,\phi^j]\\
 &\qquad+\frac12\left\{
 \Psi_L^tC\rho^i[\phi^i,\Psi_L]
 +\Psi_L^\dagger C(\rho^i)^*[\phi^i,\Psi_L^*]
 \right\}\Big].
 \end{split}
    \label{eq:dilaton-full-expression}
\end{equation}
Here $-\frac12(D\phi^i)^2$ is replaced by $\frac12\phi^iD^2\phi^i$, while the symmetrized fermion term and all interactions are unchanged.

It is $\mathcal L'$ whose VEV appears in \eqref{eq:lagrangian} and \eqref{eq:cft4d}. It obeys $\mathcal L'=(1-\gamma\varepsilon(x^3))\mathcal L'_{\rm ref}$, where $\mathcal L'_{\rm ref}=\bar g^{-2}\mathcal L'_{0SYM}$ is the perturbing operator in section \ref{sec:review}. The local factor $g(x^3)^{-2}$ cancels $g_\pm^2$ from the gauge propagator.

The interface terms in \eqref{eq:interfaceL} do not change the identification of the ambient operator $\mathcal{L}'$. They change the boundary conditions of the fields, especially for the fermion fields since they are bilinear in the Lagrangian. Therefore, nontrivial calculations are required for their contributions to the 1-pt function. We derived the boundary conditions for the fermions for different SUSY interfaces in Appendix \ref{app1}.

\subsection{One-point function of $\mathcal{L}'$}
\label{sec:4d-onepoint}
We evaluate the bulk one-point function at leading order in the weak coupling limit. The subscript $\mathrm{fin}$ below denotes subtraction of the vacuum contribution at the local coupling $g(x^3)$.

As mentioned above, although the expression for $\mathcal{L}'$ remains the same as in Section \ref{sec:review}, the fermions obtain new interface boundary conditions and could have nontrivial effects. Concretely, their contributions are obtained by differentiating the normal propagator $\langle\psi(x)\bar\psi(y)\rangle$ and taking the coincidence limit. We performed the calculation in Appendix \ref{app2} for the fermion propagators under the new boundary conditions obtained in \ref{app1}. The results are that the normal propagator remains the ordinary vacuum one at the local coupling when both points lie on the same side, for every fermion species. The interface bilinears produce mirror reflection terms due to the interface terms only in the anomalous propagator $\langle\psi(x)\psi(y)\rangle$, which does not enter the kinetic term. Hence we have
\begin{equation}
 \left\langle-\frac{i}{2g(x^3)^2}
 \operatorname{tr}\!\left[\bar\psi^A\bar\sigma^\mu\partial_\mu\psi^A
 -(\partial_\mu\bar\psi^A)\bar\sigma^\mu\psi^A\right]\right\rangle_{\rm fin}=0.
 \label{eq:freefermion}
\end{equation}

The scalar insertion $\mathrm{tr}(\phi^i\partial^2\phi^i)$ likewise has no finite bulk contribution by the free field equation since its boundary condition remains the same. Thus only the gauge-field image term remains \cite{Clark_2005}, giving
\begin{equation}
 \langle\mathcal{L}'(x)\rangle
 =-\frac{1}{4g(x^3)^2}\langle\mathrm{tr}F_{\mu\nu}F^{\mu\nu}(x)\rangle_{\rm fin}
 =\frac{3N^2}{16\pi^2(x^3)^4}\gamma\,\varepsilon(x^3).
 \label{eq:4d-onepoint-result}
\end{equation}
This result holds for all $\mathcal N=0,1,2,4$ interfaces, to all orders in the jump $\gamma$ at leading order in the weak coupling. It agrees exactly with the $\mathcal N=4$ gravity result and is numerically consistent with exact agreement for $\mathcal N=2$. For $\mathcal N=0,1$, the weak- and strong-coupling results agree only to first order in $\gamma$.

\section{1-pt functions in 2D Janus and SUSY Janus}
\label{sec:2d}
In this section we calculate the one-point function of the dilaton dual in two Janus solutions of the D1--D5 system \cite{David_2002}: the non-supersymmetric solution \cite{Bak_2007} and the half-BPS $\mathcal{N}=(4,4)$ interface \cite{Chiodaroli_2010_half,Chiodaroli_2010}. The former has an effective three-dimensional description, while the latter is realized on $AdS_2\times S^2\times T^4\times\Sigma$. At the free-orbifold point, both are described by $(T^4)^N/S_N$, but the operators and interface conditions relevant to the comparison differ as explained below.

For the half-BPS Janus interface the strong- and weak-coupling results agree exactly. For non-supersymmetric Janus they agree only to first order in the jump parameter.

\subsection{SUSY Janus}
Let us first consider SUSY Janus \cite{Chiodaroli_2010_half}. Let us use the 6D metric in Einstein frame after integrating over $T^4$:
\begin{equation}
\begin{split}
     ds^2_{6,E}
&= \rho^2 f_3^2
\left(
\frac{\cosh^2(x+\psi)}{\cosh^2\psi\,\cosh^2\theta}
ds_{AdS_2}^2
+ dx^2 + dy^2
\right)
+ f_2^2 f_3^2\, ds^2_{S^2}\\
&=R^2~K(x,y)\left(\frac{\cosh^2(x+\psi)}{\cosh^2\psi\cosh^2\theta}ds_{AdS_2}^2+dx^2+dy^2\right)+\frac{R^2\sin^2y}{K(x,y)}ds_{S^2}^2,
\end{split}   
\label{eq:6d}
\end{equation}

where 
\begin{equation}    K(x,y)=\sqrt{1+\frac{(\cosh^2\theta\cosh^2\psi-1)\sin^2y}{\cosh^2(x+\psi)}}.
\end{equation}

Below we set $R=1$ and consider the supergravity case where only the dilaton jumps across the interface and set the deformation parameter of the orbifold twist operator $\theta=0$. The 6D dilaton is
\begin{equation}    e^{-\phi(x,y)}=4\,\frac{\cosh^2(x+\psi)+\sinh^2\psi\sin^2y}{\cosh^{2}\psi\cosh^2 x}
\label{eq:10dd}
\end{equation} 

Near the two asymptotic regions $x\to\pm\infty$, its expansion is

\begin{equation}
\begin{split}
        \phi(x,y)=\phi_\pm
        +2\left(1-e^{\mp 2\psi}(\cos 2y+2\cosh^2\psi\sin^2y)\right)e^{\mp 2x}+\dots
\end{split}
\label{eq:expand6d}
\end{equation}

Notice that the 6D dilaton $\phi=\phi_6$ here is related to the 10D dilaton $\phi_{10}$ by $e^{-\phi_6}=e^{-\phi_{10}}f_3^4$, where the $f_3^4$ metric factor also contains a copy of $e^{-\phi_{10}}$. Hence the relation between the 6D dilaton asymptotic values $\phi_\pm$ and the Janus parameter $\psi$ is incorporated in a jump parameter $\gamma$ via
\begin{equation}
    \gamma  = \tanh\left(\frac{\phi_--\phi_+}{4}\right) = \tanh \psi,
\end{equation}
where the definition of $\gamma$ here has an extra factor of 2 when compared to the definition of the analogous 4d quantity. To extract the 1-pt function using AdS/CFT dictionary, one needs to go to the 3D 'effective model' of super-Janus and calculate the effective 3D dilaton $\Phi$ in the Einstein-Hilbert action \cite{baig2024transmissioncoefficientsuperjanussolution}. Namely, in the process of KK reducing the 6D action from \eqref{eq:6d} to an effective 3D action, $\Phi$ is further shifted by the position-dependent Newton's constant $G(x,y)$. Integrating out the internal $S^2$ and $y$ direction, the 3D dilaton is then
\begin{equation}
    \Phi(x)=\frac{1}{V_{S^3}}\int dy\,dV_{S^2}\big(\phi(x,y)-\log K(x,y)\big)
\end{equation}

The expansion of the 3D dilaton is 
\begin{equation}
    \Phi(x,y)=\Phi_\pm
        +2\left(1-e^{\mp 2\psi}\cosh2\psi\right)e^{\mp 2x}+\dots
\end{equation}

The AdS$_2$ metric factor in the 6D solution (or the effective 3D warpfactor) near the asymptotic regions is
\begin{equation}
    \lim_{x\to\pm\infty}f_{AdS_2}^{2}=\frac{e^{\pm 2\psi}}{4\cosh^2\psi}e^{\pm 2x},
\end{equation}

where $f=zf_{AdS_2}^{-1}$ is the appropriate Fefferman--Graham parameter near the boundary. At the asymptotic boundary, $z=|x^1|$, while the two signs of the slicing coordinate $x$ label the two signs of $x^1$. Expanding the 3D dilaton in terms of $f$ gives
\begin{equation}
    e^{\Phi(x)}=e^{\Phi_{\pm}}(1+b_f\,f^{2}+\dots),
\end{equation}

where
\begin{equation}
    b_f^\pm=\frac{e^{\pm 2\psi}}{2z^2\cosh^2\psi}\left(1-e^{\mp 2\psi}\cosh2\psi\right).
\end{equation}

Recall that we can read off the Lagrangian VEV from expansion of dilaton. From AdS$_3$/CFT$_2$ dictionary, if an action is deformed by a massless primary field with dimension $\Delta=2$ coupled to a source, the expansion of the bulk field $\phi_\Delta$ near the boundary is 
\begin{equation}
     \phi_{\Delta=2}\sim a_f(1+\dots)+b_f(f^2+\dots)
 \end{equation}
 
 In the dual field theory, the coupling of the field is $a_f$, while the VEV of the field is $\langle O_{\Delta=2}\rangle=-2b_f$. The dual operator $O$ is related to the Lagrangian by a factor of $1/16\pi G_3$ \cite{Clark_2005,aharony2024typeiistringtheory}. Using the Brown-Henneaux relation $c=\tfrac{3}{2G_3}$ and $c=6N=6Q_1Q_5$ in the D1-D5 CFT, the 1-pt function of the Lagrangian is
\begin{equation}
\begin{split}
    \langle\mathcal{L}\rangle_{\mathrm{grav}}&=\frac{-2}{16\pi G_3}(b_f^+\theta(x)+b_f^-\theta(-x))\\
    &=-\frac{N}{4\pi z^2}\frac{e^{2\psi}-\cosh2\psi}{\cosh^2\psi}\theta(x)+\frac{N}{4\pi z^2}\frac{e^{-2\psi}-\cosh2\psi}{\cosh^2\psi}\theta(-x)=\frac{N\gamma}{2\pi(x^1)^2}\varepsilon(x^1)   
\end{split}    
\label{eq:2dsujanus}
\end{equation}

\subsection{The free orbifold CFT}
In 2D the dilaton is dual to an untwisted $(h,\bar h)=(1,1)$ modulus of the $\mathcal{N}=(4,4)$ theory. For the half-BPS interface, supersymmetry completes the bosonic radius deformation by fermionic terms, schematically $\gamma\varepsilon(x^1)[(\partial X)^2+\Psi\partial\Psi^\dagger+\dots]$. We denote this full marginal operator by $\mathcal{L}$. The non-supersymmetric dilatonic solution considered later instead couples to the bosonic current-current operator $O_\phi$ \cite{Bak_2007}.

The 2D $\mathcal{N}=(4,4)$ orbifold CFT dual to the super-Janus solution is a free field theory with target space $(T^4)^N/S_N$ where $N=Q_1Q_5$. The radii of boson and fermion jump across the interface. We only consider half-BPS interfaces in the case of super-Janus. We can write down the continuous action of the free field theory on one $T^4$ in terms of complex fields \cite{David_2002,Chiodaroli_2010}:
\begin{equation}
\begin{split}
        S&=\frac{1}{2}\int d\tau d\sigma\bigg(\frac{1}{2}(\partial_-X^I\partial_+X^{I\dagger}+\partial_+X^I\partial_-X^{I\dagger})\\
    &-(\Psi^I\partial_-\Psi^{I\dagger}-\frac{1}{2}\partial_-(\Psi\Psi^{I\dagger})+\tilde{\Psi}^I\partial_+\tilde{\Psi}^{I\dagger}-\frac{1}{2}\partial_+(\tilde{\Psi}\tilde{\Psi}^{I\dagger})\bigg).
\end{split}
\end{equation}

Here, $I\in\{1,2\}$ labels the two complex boson/fermion fields in each copy of $T^4$, and we suppress the copy index $1,\dots,N$. Assuming that every boson has the same radius, $r_i^I=r_i$, we define the radius-squared (equivalently, isotropic sigma-model-metric) jump by
\begin{equation}
     \frac{r_1^2-r_2^2}{r_1^2+r_2^2} = \tanh\psi=\gamma
\end{equation}
This is not literally the relative $T^4$-volume jump: for $V_{T^4}\propto r^4$, the latter is $2\gamma/(1+\gamma^2)$.

The convention here is that $r_1$ corresponds to the $x\to\infty$ region in gravity , and $r_2$ is $x\to -\infty$.

The interface sitting at $\sigma=0$ is where the radii of the bosons and fermions jump from $r_1$ to $r_2$. It is shown that for fixed CFT$_1$ and CFT$_2$ on the moduli space, there are two interface that preserves half of the $\mathcal{N}=(4,4)$. They correspond to the analogous type A and type B SUSY boundaries for $\mathcal{N}=(2,2)$ theories, by rotating the moduli space to either the $(cc)$ or $(aa)$ deformation moduli via the $SO(4)_{\mathrm{ext}}$ automorphism coming from the superalgebra of $\mathcal{N}=4$. In the present case the deformation is in the volume of $T^4$ in the sigma model, which is the K\"ahler moduli, so the only interface that preserves the supersymmetry is a type B interface. The boundary conditions for bosons at the interface $\sigma=0$ are:
\begin{equation}
    \frac{X_1^I}{r_1}=\frac{X_2^I}{r_2},\quad r_1\partial_\sigma X_1^I=r_2\partial_\sigma X_2^I.
    \label{eq:bosbc}
\end{equation}

This determines the complex boson propagator from the mirror charge method \cite{Azeyanagi_2008}
\begin{equation}
\begin{split}
        \langle X_1^I (x,\bar{x}) X_1^{J\dagger}(y,\bar{y})\rangle = -\frac{\delta^{IJ}}{\pi}\left(\log|x-y|^2-\gamma \log|x-\bar{y}|^2\right)\\
        \langle X_2^I (x,\bar{x}) X_2^{J\dagger}(y,\bar{y})\rangle = -\frac{\delta^{IJ}}{\pi}\left(\log|x-y|^2+\gamma \log|x-\bar{y}|^2\right)
\end{split}
\end{equation}

The fermion boundary condition is determined by the variance of the action under supercharges $G^1,G^2$ and $\tilde{G}^1,\tilde{G}^2$:
\begin{equation}
\begin{split}
        \delta_1 S=\int\epsilon_1(\Psi^1\partial_-X^2-\Psi^{2\dagger}\partial_-X^1),\quad \delta_{\tilde{1}} S=\int\tilde{\epsilon}_1(\tilde{\Psi}^1\partial_+X^2-\tilde{\Psi}^{2\dagger}\partial_+X^1)\\
    \delta_2 S=\int\epsilon_2(\Psi^2\partial_-X^2-\Psi^{1\dagger}\partial_-X^1),\quad \delta_{\tilde{2}} S=\int\tilde{\epsilon}_2(\tilde{\Psi}^1\partial_+X^2-\tilde{\Psi}^{2\dagger}\partial_+X^1).
\end{split}
\end{equation}

A Type B interface corresponds to $G^1-\tilde{G}^1=0$, $G^2-\tilde{G}^2=0$ and their conjugates. This gives fermion boundary condition \cite{Chiodaroli_2010}: 
\begin{equation}
    r_1(\Psi_1^I-\tilde{\Psi}_1^I)=r_2(\Psi_2^I-\tilde{\Psi}_2^I),\quad \frac{\Psi_1^I+\tilde{\Psi}_1^I}{r_1}=\frac{\Psi_2^I+\tilde{\Psi}_2^I}{r_2}.
    \label{eq:fermbc}
\end{equation}

A mutually orthogonal set of solutions for the free fermion field with momentum $p$ is 
\begin{equation}
\begin{split}
        \Psi_{1p}^I=\alpha_1e^{ip(\tau+\sigma)}+\beta_1e^{-ip(\tau+\sigma)},\quad \Psi_{2p}^I=\alpha_2e^{ip(\tau+\sigma)}+\beta_2e^{-ip(\tau+\sigma)}\\
    \tilde{\Psi}_{1p}^I=\tilde{\alpha}_1e^{ip(\tau-\sigma)}+\tilde{\beta}_1e^{-ip(\tau-\sigma)},\quad \tilde{\Psi}_{2p}^I=\tilde{\alpha}_2e^{ip(\tau-\sigma)}+\tilde{\beta}_2e^{-ip(\tau-\sigma)},\\
\end{split}
\end{equation}

which satisfy the proper normalization conditions $\alpha_i^2+\beta_i^2=\tilde{\alpha}_i^2+\tilde{\beta}_i^2=1$. For simplicity, we only consider solutions that identify the left-moving and right-moving ones $\alpha_i=\tilde{\alpha}_i, \beta_i=\tilde{\beta}_i$. In addition, we impose that $\alpha_1\beta_2+\alpha_2\beta_1=0$, so that the correlator across the interface $\langle\Psi_1\Psi_2^{\dagger}\rangle$ only has one component. The first boundary condition in \eqref{eq:fermbc} on the interface $\sigma=0$ is automatically satisfied, while the second one demands that $(\alpha_1+\beta_1)/r_1=(\alpha_2+\beta_2)/r_2$. Coefficients that satisfy that demand are
\begin{equation}
    \alpha_1=\alpha_2=\frac{r_1+r_2}{\sqrt{2(r_1^2+r_2^2)}}, \beta_1=-\beta_2=\frac{r_1-r_2}{\sqrt{2(r_1^2+r_2^2)}}.
\end{equation}

The propagators of the 2D complex fermions are then
\begin{equation}
\begin{split}
    \langle\Psi_1^I(x)\Psi_1^{J\dagger}(y)\rangle=\frac{\delta^{IJ}}{\pi}\left(\frac{1}{x-y}-\gamma\frac{1}{x-\bar{y}}\right)\\
    \langle\Psi_2^I(x)\Psi_2^{J\dagger}(y)\rangle=\frac{\delta^{IJ}}{\pi}\left(\frac{1}{x-y}+\gamma\frac{1}{x-\bar{y}}\right).
\end{split}
\end{equation}

The one-point function of the Lagrangian is extracted by applying $\partial_-\partial_+$ to the boson propagators and $\partial_\mp$ to the chiral and anti-chiral fermion propagators before taking $x=y$. The only non-divergent contributions come from the image terms:
\begin{equation}
    \langle \mathcal{L}\rangle_{\mathrm{free}}
    =\frac{2N}{\pi}\,\gamma\,\frac{1}{(2x^1)^2}\varepsilon(x^1)
    =\frac{N\gamma}{2\pi(x^1)^2}\varepsilon(x^1).
    \label{eq:cft}
\end{equation}

Compared to \eqref{eq:2dsujanus}, the CFT and gravity one-point functions match exactly for the supersymmetry-preserving interface.

\subsection{Non-supersymmetric Janus}
We now consider the supersymmetry-breaking interface \cite{Bak_2007,Azeyanagi_2008}. This is a top-down type-IIB solution with an effective three-dimensional Einstein-frame description. Let $y$ denote the bulk slicing coordinate, set the AdS radius to one, and define
\begin{equation}
 s\equiv\sqrt{1-2b^2},\qquad |b|<\frac{1}{\sqrt{2}}.
\end{equation}
The metric and dilaton are
\begin{align}
 ds_3^2&=dy^2+F(y)ds_{AdS_2}^2,
 &F(y)&=\frac{1}{2}\left(1+s\cosh 2y\right),\\
 \phi(y)&=\phi_0+\frac{1}{\sqrt{2}}
 \log\!\left(\frac{1+s+\sqrt{2}\,b\tanh y}
 {1+s-\sqrt{2}\,b\tanh y}\right).
\end{align}
Their asymptotic behavior at $y\to\pm\infty$ is
\begin{align}
 \phi_\pm&=\phi_0\pm\frac{1}{\sqrt{2}}
 \operatorname{arctanh}(\sqrt{2}b),
 &F(y)&\sim\frac{s}{4}e^{\pm2y},\\
 \phi(y)&=\phi_\pm\mp\frac{2b}{s}e^{\mp2y}+\dots
\end{align}
Writing $ds_{AdS_2}^2=(dz^2-dt^2)/z^2$ and introducing the Fefferman--Graham coordinate $\rho=z/\sqrt{F(y)}$, the last expansion becomes
\begin{equation}
 \phi=\phi_\pm\mp\frac{b}{2z^2}\rho^2+\dots .
\end{equation}
Consequently, with $x_\perp\equiv x^1$ the distance from the interface, the gravity one-point function is \cite{Bak_2007}
\begin{equation}
 \langle \mathcal{L}\rangle_{\mathrm{grav}}
 =\frac{Nb}{4\pi x_\perp^2}\varepsilon(x_\perp).
 \label{eq:2djanus}
\end{equation}
Here and below we use the sign convention $\mathcal{L}=-O_\phi$ relative to the operator $O_\phi$ in \cite{Bak_2007}. Exchanging the two sides flips the signed result.

Two distinctions from the half-BPS calculation are important. First, the compactification radius that enters the canonically normalized free CFT obeys $r^2\propto e^{-\phi}$, or $r\propto e^{-\phi/2}$; the Einstein-frame geometric length $e^{-\phi/4}$ is not the CFT radius \cite{Azeyanagi_2008}. Choosing the side labels so that $b$ and $\gamma$ have the same sign therefore gives
\begin{equation}
\begin{split}
 \gamma&=\frac{r_1^2-r_2^2}{r_1^2+r_2^2}
 =\tanh\!\left[\frac{1}{\sqrt{2}}\operatorname{arctanh}(\sqrt{2}b)\right]\\
 &=\frac{(1+\sqrt{2}b)^{1/\sqrt{2}}-(1-\sqrt{2}b)^{1/\sqrt{2}}}
 {(1+\sqrt{2}b)^{1/\sqrt{2}}+(1-\sqrt{2}b)^{1/\sqrt{2}}}
 =b+\frac{b^3}{3}+O(b^5).
\end{split}
\end{equation}
Thus $b=\gamma-\gamma^3/3+O(\gamma^5)$. Then we have the gravity result for the non-BPS Janus interface
\begin{equation}
 \langle\mathcal{L}\rangle_{\mathrm{grav}}
 =\frac{N}{4\pi x_\perp^2}
 \left(\gamma-\frac{\gamma^3}{3}+O(\gamma^5)\right)\varepsilon(x_\perp).
\end{equation}

On the other hand, the nonsupersymmetric dilaton operator is the bosonic current operator, and the fermionic contribution in the previous section drops out. The free CFT result is therefore
\begin{align}
 \langle\mathcal{L}\rangle_{\mathrm{free}}
 =\frac{N\gamma}{4\pi x_\perp^2}\varepsilon(x_\perp)
\end{align}
They agree at linear order, but the first mismatch occurs at $O(\gamma^3)$.

\section{Conclusion and future directions}

In this paper we studied one-point functions of marginal operators dual to the space-varying dilaton in supersymmetric and non-supersymmetric Janus interfaces in 4D and 2D. On the gravity side, we extracted the one-point function from the asymptotic dilaton expansion. On the field-theory side, we computed it in weakly coupled 4D $\mathcal{N}=4$ SYM with a gauge-coupling jump and at the free-orbifold point of the 2D D1--D5 CFT with a jump in the $T^4$ sigma-model modulus.

In 4D, the weak- and strong-coupling results agree only to first order in $\gamma$ for the non-SUSY and $\mathcal N=1$ interfaces. The $\mathcal N=4$ result agrees exactly, while numerical evaluation of the full $\mathcal N=2$ integral is consistent with exact agreement. All four cases share the protected linear term expected from conformal perturbation theory \cite{Clark_2005}. In 2D, exact agreement holds for the half-BPS interface. For non-SUSY Janus, the results agree at leading order and first differ at $O(\gamma^3)$.

The 4D dilaton operator is given by \eqref{eq:dilaton-full-expression}. Only its scalar kinetic term differs from the usual SYM density. The interface bilinears generate anomalous fermion reflection, but the normal propagator has no image term. The symmetrized kinetic insertion and the fermion-current derivative therefore have separately vanishing finite VEVs. This holds for every species at Gaussian order and to all orders in the jump, giving the same weak-coupling result for all $\mathcal N=0,1,2,4$ interfaces.

There are several directions for future work. First, it would be useful to derive analytically the simple relation inferred from the full $\mathcal{N}=2$ integral and to identify the protection mechanism behind the exactly matching cases for eight supercharges. Conformal perturbation theory \cite{Clark_2005} and supersymmetric localization \cite{He_2025} may provide useful approaches. Testing the weak/strong-coupling match for more general D3/D5/NS5 defects would also be interesting \cite{He_2025,linardopoulos2025stringtheorymethodsdefect}.

Second, in 2D it would be interesting to look for holographic interfaces preserving less supersymmetry. At present, the examples considered in Section \ref{sec:2d} are the non-supersymmetric and half-BPS Janus interfaces. The 4D $\mathcal{N}=2$ result suggests that exact matching of this observable may survive with less than half of the supersymmetry; constructing intermediate 2D examples would test whether an analogous mechanism exists there. One can also ask whether similar phenomena occur for holographic interfaces in other dimensions.

Finally, it would be worthwhile to understand other ICFT or defect CFT observables for holographic interfaces. The 1-pt function $a_O$ is among the simplest pieces of ICFT data, but it is closely tied to other data like the bulk-to-defect correlators \cite{Bill__2016}. It would be interesting to study the behavior of BOPE coefficients in Janus interfaces, especially those involving the displacement operator. 

\section*{Acknowledgments}
We'd like to thank Christoph Uhlemann for useful discussions and for pointing out mistakes in the first version of this manuscript. Some of the corrections made in response to \cite{christophtoappear} were done with the help of OpenAI's \texttt{gpt-5.6-sol} and \texttt{gpt-6 Astra}. This work was supported in part by DOE grant DE-SC0022021 and by a grant from the Simons Foundation (Grant 651678, AK).

\bibliographystyle{JHEP}
\bibliography{references_v2}

\clearpage
\appendix
\section{The fermion boundary conditions}
\label{app1}
In this appendix we derive the fermion boundary conditions from the quadratic terms in \eqref{eq:symlag} and \eqref{eq:interfaceL}. Following \cite{D_Hoker_2006}, we vary the action for a smooth, positive coupling $g(x^3)$ before taking the discrete coupling jump at the interface. 

We work in weak coupling limit and suppress gauge and species indices. The interface bilinear couples to all four Weyl fermions for the $\mathcal{N}=4$ interface, and to two and one species for $\mathcal{N}=2$ and $\mathcal{N}=1$, respectively, as selected by $D^{(2)}$ and $D^{(1)}$ in \eqref{eq:interfaceL}. We first consider a selected species, then the remaining fermions.

\subsection{The smooth quadratic action}
We use the conventions
\begin{equation}
 \gamma^\mu=\begin{pmatrix}0&\sigma^\mu\\ \bar\sigma^\mu&0\end{pmatrix},
 \qquad \sigma^\mu=(\mathbf 1,\vec\sigma),
 \qquad \bar\sigma^\mu=(-\mathbf 1,\vec\sigma).
 \label{eq:app-spinor-conventions}
\end{equation}
Writing $\varepsilon=i\sigma^2$, a compatible charge-conjugation matrix is
\begin{equation}
 C=-\gamma^3\gamma^1=
 \begin{pmatrix}-\varepsilon&0\\0&-\varepsilon\end{pmatrix},
\end{equation}
We take $\varepsilon_{\alpha\beta}=(i\sigma^2)_{\alpha\beta}$ and $\varepsilon^{\alpha\beta}=(-i\sigma^2)^{\alpha\beta}$, with the analogous convention for dotted indices. For $\Psi_L=(\psi,0)^t$, the bilinears in \eqref{eq:interfaceL} are then $\Psi_L^tC\Psi_L=-\psi\psi$ and $\Psi_L^\dagger C\Psi_L^*=\bar\psi\bar\psi$.

The quadratic bulk and interface terms for one selected Weyl fermion are ($g'=\partial_3g$)
\begin{equation}
 \mathcal L_B+\mathcal L_I
 =-\frac{i}{2g^2}\left[
 \bar\psi\bar\sigma^\mu\partial_\mu\psi
 -(\partial_\mu\bar\psi)\bar\sigma^\mu\psi\right]
 +\frac{ig'}{2g^3}\left(-\psi\psi+\bar\psi\bar\psi\right).
 \label{eq:app-smooth-action}
\end{equation}
 Define
\begin{equation}
 q(x^3)=g(x^3)^{-2},\qquad q_\pm=g_\pm^{-2},
 \qquad \bar q=\bar g^{-2}=\frac{q_++q_-}{2}
 \end{equation}

 and the jump parameters
 \begin{equation}
 \eta=\log\frac{g_+}{g_-},\qquad
 \gamma=\frac{g_+^2-g_-^2}{g_+^2+g_-^2}=\tanh\eta,
 \label{eq:app-couplings}
\end{equation}
where $g_\pm>0$ and $\gamma$ is the signed jump in \eqref{eq:gamma}.

Varying the smooth bulk action $S_B=\int d^4x\,\mathcal L_B$ with respect to $\bar\psi$ and integrating by parts gives
\begin{equation}
 \frac{\delta S_B}{\delta\bar\psi}
 =-i\left(q\bar\sigma^\mu\partial_\mu\psi
 +\frac{q'}{2}\bar\sigma^3\psi\right).
 \label{eq:app-smooth-variation}
\end{equation}
Note that we have an additional term of the derivative of $q$ in terms of the space-varying coupling.

Consider the canonically normalized Weyl field $\lambda$ corresponding to the fermion rescaling in section \ref{sec:review} given by $\psi=g\lambda$.  
Set $h(x^3)=\partial_3\log g=-\frac{q'}{2q}$.
The derivatives of $g$ cancel between the two symmetrized kinetic terms, so that
\begin{equation}
 \mathcal L_B+\mathcal L_I
 =-\frac{i}{2}\left[
 \bar\lambda\bar\sigma^\mu\partial_\mu\lambda
 -(\partial_\mu\bar\lambda)\bar\sigma^\mu\lambda\right]
 +\frac{ih}{2}\left(-\lambda\lambda+\bar\lambda\bar\lambda\right).
 \label{eq:app-canonical-action}
\end{equation}

Before taking $g$ to be the discrete jump limit, we fix the conventions of the Weyl fermions. Let $b$ be the column of lower dotted components of $\bar\lambda$. We fix the charge-conjugate phase by
\begin{equation}
 \lambda\lambda=\lambda^t\varepsilon\lambda,
 \qquad \bar\lambda\bar\lambda=-b^t\varepsilon b,
 \qquad H_\lambda=-\sigma^3\varepsilon b.
 \label{eq:app-charge-conjugate}
\end{equation}
After Fourier transforming parallel to the interface with dependence $e^{ip_ax^a}$, $a=0,1,2$, define
\begin{equation}
 \mathsf A(p)=-i\bar\sigma^3p_a\bar\sigma^a,
 \qquad \Upsilon_\lambda=
 \begin{pmatrix}\lambda\\ H_\lambda\end{pmatrix}.
\end{equation}
The EOM for fermions obtained from \eqref{eq:app-canonical-action} take the form
\begin{equation}
 \partial_3\Upsilon_\lambda
 =\left[
 \begin{pmatrix}\mathsf A&0\\0&-\mathsf A\end{pmatrix}
 +h(x^3)\begin{pmatrix}0&\mathbf 1\\\mathbf 1&0\end{pmatrix}
 \right]\Upsilon_\lambda.
 \label{eq:app-smooth-nambu}
\end{equation}
Each block acts on the two Weyl components. The rephasing $H_\lambda\to-H_\lambda$ reverses the off-diagonal entries and the anomalous amplitudes below. We use \eqref{eq:app-charge-conjugate} throughout.

\subsection{The discrete jump limit}
Consider the Janus coupling with width $\ell$. We now take $\ell\to0$ and $g(x^3)$ interpolates between $g_-$ and $g_+$. The contribution from the first term in \eqref{eq:app-smooth-nambu} then vanishes, while
\begin{equation}
 \lim_{\ell\to0}\int\,dx^3\,h(x^3)
 =\log\frac{g_+}{g_-}=\eta.
\end{equation}
Consider the fermion transfer matrix from the left side to the right
\begin{equation}
 \Upsilon_{\lambda,+}=\mathsf M(\eta)\Upsilon_{\lambda,-},
 \qquad
 \mathsf M(\eta)=
 \begin{pmatrix}
 \cosh\eta\,\mathbf 1&\sinh\eta\,\mathbf 1\\
 \sinh\eta\,\mathbf 1&\cosh\eta\,\mathbf 1
 \end{pmatrix}.
 \label{eq:app-transfer}
\end{equation}

Now we can write down the boundary conditions for the fermion at the interface. Consider $\psi_\pm$ on left and right side of the interface, and denote $H=gH_\lambda=-\sigma^3\varepsilon\bar\psi_{\rm col}$, where $\bar\psi_{\rm col}$ is the column of lower dotted components of $\bar\psi$. The combinations $\lambda\pm H_\lambda$ have transfer factors $e^{\pm\eta}$. Multiplying by $g_+/g_-=e^\eta$ gives the boundary condition
\begin{equation}
 \begin{split}
 \psi_+-H_+&=\psi_--H_-,\\
 q_+(\psi_++H_+)&=q_-(\psi_-+H_-).
 \end{split}
 \label{eq:bcfermionint}
\end{equation}

The above boundary condition applies to the species selected by the bilinear term in the interface actions. $\psi$ mixes with part of $\bar{\psi}$. For the fermion species that does not appear in the interface action (for $\mathcal{N}=1,2$ interfaces), EOM of \eqref{eq:app-smooth-variation} gives
\[
 q\bar\sigma^\mu\partial_\mu\psi
 +\frac{q'}{2}\bar\sigma^3\psi=0.
\]
The kinetic coefficient still varies across the interface. Using $q'/(2q)=-g'/g$ and $\psi=g\lambda$, with $g=g(x^3)$, we find
\[
 \begin{aligned}
 0&=\bar\sigma^\mu\partial_\mu(g\lambda)
    -\frac{g'}{g}\bar\sigma^3(g\lambda)\\
  &=g\bar\sigma^\mu\partial_\mu\lambda
    +g'\bar\sigma^3\lambda-g'\bar\sigma^3\lambda
   =g\bar\sigma^\mu\partial_\mu\lambda.
 \end{aligned}
\]
Thus $\lambda$ obeys the free Weyl equation throughout the transition region $\ell$. Fourier transforming as in \eqref{eq:app-smooth-nambu} gives $\partial_3\lambda=\mathsf A(p)\lambda$. In the original fields, the interface condition for those absent in the interface term is
\begin{equation}
 \frac{\psi_+}{g_+}=\frac{\psi_-}{g_-},
 \qquad
 \frac{\bar\psi_+}{g_+}=\frac{\bar\psi_-}{g_-}.
 \label{eq:app-unmixed-matching}
\end{equation}

\section{Fermion propagators and the one-point function}
\label{app2}
In this appendix we derive the propagators using the above boundary conditions and the vacuum-subtracted fermion insertions used in section \ref{sec:4d-onepoint}. We solve the matching conditions \eqref{eq:bcfermionint} for species coupled to the interface bilinear, and \eqref{eq:app-unmixed-matching} for the remaining species. Reflection occurs in the anomalous channel $\langle\psi\psi\rangle$, while the normal channel $\langle\psi\bar\psi\rangle$ determines the kinetic one-point function.

\subsection{Normal and anomalous propagators}
 For one Weyl species, the time-ordered propagators are
\begin{equation}
 \begin{split}
 G^{ab}_{\alpha\dot\beta}(x,y)
 &=\langle\psi^a_\alpha(x)\bar\psi^b_{\dot\beta}(y)\rangle,\\
 F^{ab}_{\alpha\beta}(x,y)
 &=\langle\psi^a_\alpha(x)\psi^b_\beta(y)\rangle,
 \qquad a,b\in\{+,-\}.
 \end{split}
 \label{eq:app-correlators}
\end{equation}
The labels $a,b$ specify the half-spaces containing $x,y$, respectively, and $\alpha,\dot\beta$ are two-component spinor indices. Below we suppress spinor indices in matrix notation. The convention is that $t_+$ and $t_-$ are the coefficients of $G^{-+}$ and $G^{+-}$.

Let $X=x-y$ and $\widetilde X=x-Ry$, where $Ry=(y^0,y^1,y^2,-y^3)$. We use the spinor matrices in \eqref{eq:app-spinor-conventions} and $\varepsilon=i\sigma^2$. The free Weyl kernel with unit kinetic normalization is
\begin{equation}
 G^{(0)}(X)=-i\sigma^\mu\partial_{X^\mu}
 \frac{1}{4\pi^2(X^2-i0)}
 =\frac{iX_\mu\sigma^\mu}{2\pi^2(X^2-i0)^2}.
 \label{eq:app-free-weyl}
\end{equation}
For a selected species, the normal propagators are
\begin{equation}
 \begin{aligned}
 G^{++}(x,y)&=g_+^2G^{(0)}(X),
 &G^{--}(x,y)&=g_-^2G^{(0)}(X),\\
 G^{-+}(x,y)&=t_+G^{(0)}(X),
 &G^{+-}(x,y)&=t_-G^{(0)}(X).
 \end{aligned}
 \label{eq:propagator}
\end{equation}

The full normal propagator is the sum of these blocks with the corresponding half-space step functions.

On the other hand, due to the interface boundary condition we also have anomalous propagators
\begin{equation}
 F^{\pm\pm}(x,y)
 =\frac{ig_\pm^2s_\pm}{2\pi^2}
 \frac{\widetilde X_\mu\sigma^\mu\bar\sigma^3\varepsilon}
 {(\widetilde X^2-i0)^2},
 \qquad s_+=-\gamma,\qquad s_-=\gamma,
 \label{eq:app-anomalous-propagator}
\end{equation}
whereas $F^{-+}=F^{+-}=0$. The barred anomalous propagators follow by fermionic exchange and conjugation with the index conventions in \eqref{eq:app-charge-conjugate}. In particular, $F^{ab}(x,y)=-[F^{ba}(y,x)]^t$. Under endpoint exchange, the reflected argument becomes $y-Rx=-R\widetilde X$.

Imposing the boundary condition \eqref{eq:bcfermionint} with a source on the $+$ side for the fermion species present in the interface action, we have
\begin{equation}
 t_+=t_-
 =\frac{2g_+^2g_-^2}{g_+^2+g_-^2}\equiv\bar g^2.
 \label{eq:app-normal-transmission}
\end{equation}
and
\begin{equation}
 t_+=g_+^2(1+s_+)=g_-^2(1-s_+),
\end{equation} 

Denoting the coefficients of the normal images $g_\pm^2G^{(0)}(\widetilde X)$ by $r_\pm$, we obtain\footnote{In the previous version of this manuscript we had $r_\pm\neq0$. Here we point out that mistake. The bulk Weyl equations itself provide an independent check that this normal image propagator vanishes. Away from the source and interface, we have $ \bar\sigma^\mu\partial_{x^\mu}G(x,y)=0$ and $ (\partial_{y^\mu}G(x,y))\bar\sigma^\mu=0$.
Consider the naive image propagator $G^{(0)}(\widetilde X)=\widetilde X_\nu\sigma^\nu/(\widetilde X^2)^2$ for $\widetilde X=(0,0,0,d)$ with $d=2x^3\ne0$. Its derivative is
$\left.\partial_{\widetilde X^\mu}\,G^{(0)}(\widetilde X)\right|_{v=(0,0,0,d)}
 =\frac{1}{d^4}(-\mathbf1,\sigma^1,\sigma^2,-3\sigma^3)_\mu$.

Since $\partial_{y^\mu}=-R^\nu{}_{\mu}\partial_{v^\nu}$ on this image, the contact condition for it is
\begin{equation}
 \left.\bar\sigma^\mu\partial_{x^\mu}G^{(0)}(\widetilde X)\right|_{y=x}=0,
 \qquad
 \left.(\partial_{y^\mu}G^{(0)}(\widetilde X))\bar\sigma^\mu\right|_{y=x}
 =-\frac{6}{(2x^3)^4}\mathbf1.
 \label{eq:app-image-endpoint-test}
\end{equation}
The normal image propagator then fails the second boundary condition. The correct image term is instead the anomalous tensor in \eqref{eq:app-anomalous-propagator} that includes the reflected spinor transformation and satisfies the same-chirality field equations. Hence $r_\pm=0$.}.
\begin{equation}
 r_\pm=0,
 \qquad s_+=-\gamma,\qquad s_-=\gamma.
 \label{eq:correct-image-coefficients}
\end{equation}

For an unmixed species, the same-side normal propagators are those in \eqref{eq:propagator}, and the anomalous ones vanish, so we have
\begin{equation}
 t_+=t_-=g_+g_-,\qquad r_\pm=s_\pm=0.
 \label{eq:app-unmixed-propagator}
\end{equation}
This follows by multiplying the free canonical propagator by $g$ at each endpoint.

\subsection{The bulk fermion 1-pt function}
Despite the nontrivial boundary conditions and nonzero anomalous propagators $F$, we will show that from the Weyl equations the 1-pt function vanishes identically for the Fermions regardless of whether it appears in the interface bilinear terms. 

For $x$ and $y$ in a neighborhood of a bulk point on side $a$, define the vacuum-subtracted normal propagator
\begin{equation}
 \Delta G(x,y)=G^{aa}(x,y)-g_a^2G^{(0)}(x-y).
\end{equation}
This subtraction uses the local kinetic normalization and removes the coincidence singularity. For $x^3\ne0$, $\Delta G$ is smooth as $y\to x$ since the delta-function sources cancel, giving
\begin{equation}
 \bar\sigma^\mu\partial_{x^\mu}\Delta G=0,
 \qquad (\partial_{y^\mu}\Delta G)\bar\sigma^\mu=0.
 \label{eq:app-subtracted-weyl}
\end{equation}
Taking spinor traces before the coincidence limit gives, separately,
\begin{equation}
 \begin{split}
 K_{\rightarrow}
 &\equiv\langle\bar\psi\bar\sigma^\mu\partial_\mu\psi\rangle_{\rm fin}=0,\\
 K_{\leftarrow}
 &\equiv\langle(\partial_\mu\bar\psi)\bar\sigma^\mu\psi\rangle_{\rm fin}=0.
 \end{split}
 \label{eq:app-kinetic-zero}
\end{equation}
For the same-side propagators above, $\Delta G$ vanishes identically. The vanishing kinetic insertions follow more generally from the massless bulk equations and the fact that $r_\pm=0$.

With $q(x^3)=g(x^3)^{-2}$ as in \eqref{eq:app-couplings}, let $\mathcal L_{SYM}^{\mathrm F}$ denote the symmetrized fermion kinetic density for the component under consideration. Then
\begin{align}
 \langle\mathcal L_{SYM}^{\mathrm F}\rangle_{\rm fin}
 &=-\frac{iq}{2}(K_{\rightarrow}-K_{\leftarrow})=0,
 \label{eq:app-sym-zero}\\
 \left\langle-\frac{iq}{2}
 \partial_\mu(\bar\psi\bar\sigma^\mu\psi)\right\rangle_{\rm fin}
 &=-\frac{iq}{2}(K_{\rightarrow}+K_{\leftarrow})=0.
 \label{eq:app-improvement-zero}
\end{align}
This gives the symmetrized kinetic result \eqref{eq:freefermion} and shows that the current-derivative contribution also vanishes separately. Obviously the anomalous propagator does not enter these $\bar\psi\psi$ bilinears, so it does not contribute. This result holds for every Weyl species, including those not selected by $D^{(2)}$ or $D^{(1)}$. It is also exact in the coupling jump at leading classical order for bulk insertions at $x^3\ne0$.

\end{document}